\documentclass[useAMS,usenatbib]{mn2e}

\usepackage{amsmath}
\usepackage{amssymb}
\usepackage{epsfig}
\usepackage{graphicx}
\usepackage{ifthen}
\usepackage{latexsym}
\usepackage{rotating}
\usepackage{subfigure}
\usepackage{times,epsf}
\usepackage{txfonts}
\usepackage{varioref}
\usepackage{verbatim}
\usepackage{url}
\usepackage{color}
\usepackage{array}
\usepackage[dvipsnames]{xcolor}
\usepackage[T1]{fontenc}

\newcolumntype{M}{>{$\vcenter\bgroup\hbox\bgroup}c<{\egroup\egroup$}}

\newcommand{\be}{\begin{equation}}
\newcommand{\ee}{\end{equation}}
\newcommand{\bea}{\begin{eqnarray}}
\newcommand{\eea}{\end{eqnarray}}

\def\gsim{ \lower .75ex \hbox{$\sim$} \llap{\raise .27ex \hbox{$>$}} }
\def\lsim{ \lower .75ex\hbox{$\sim$} \llap{\raise .27ex \hbox{$<$}} }

\newcommand\actaa{Acta Astronomica}

\newcommand\apj{Astrophysical Journal}
\newcommand\apjl{Astrophysical Journal Letters}

\newcommand\apss{Astrophysics and Space Sciences}

\newcommand\aap{Astronomy \& Astrophysics}

\newcommand\nat{Nature}

\newcommand\mnras{Monthly Notices of the Royal Astronomical Society}

\newcommand\pasj{Publications of the Astronomical Society of Japan}

\newcommand{\Medd}{\dot M_{\rm Edd}}

\title[Jet and disk luminosities]{Jet and disk luminosities in tidal
disruption events}
\author[T. Piran, A. S\k{a}dowski, A. Tchekhovskoy] {Tsvi Piran$^1$\footnotemark[1], Aleksander
  S\k{a}dowski$^{2,4}$\footnotemark[1], Alexander Tchekhovskoy$^{3,4}$\thanks{E-mail: tsvi.piran@mail.huji.ac.il (TP); asadowsk@mit.edu (AS);
    atchekho@berkeley.edu (AT)} \\ $^1$ Racah Institute for Physics,
  The Hebrew University, Jerusalem 91904, Israel\\
$^2$ MIT Kavli Institute for Astrophysics and Space Research,
77 Massachusetts Ave, Cambridge, MA 02139, USA\\
$^3$ Department of Physics and Department of Astronomy, University of
California, Berkeley, CA 94720-3411\\
$^4$ Einstein Fellow }

\begin{document}

\maketitle

\label{firstpage}

\begin{abstract}
  Tidal disruption events (TDE) in which a star is devoured by a
  massive black hole at a galactic center pose a challenge to our
  understanding of accretion processes. Within a month the accretion
  rate reaches super-Eddington levels. It then drops gradually over a
  time scale of a year to sub-Eddington regimes.  The initially
  geometrically thick
  disk becomes a thin one and eventually an ADAF at very low accretion
  rates.  As such, TDEs explore the whole range of accretion rates and
  configurations. A challenging question is what the corresponding
  light curves of these events are. We explore numerically the disk
  luminosity and the conditions within the inner region of the disk using a fully
  general relativistic slim disk model.
  Those conditions determine the magnitude of the magnetic field that
  engulfs the black hole and this, in turn, determines the
  Blandford-Znajek jet power.  We estimate this power in two different ways and show that they are 
  self-consistent. We find, as expected earlier from analytic
  arguments \citep{krolik12}, that neither the disk luminosity nor the
  jet power follows the accretion rate throughout the disruption event.  The disk luminosity
   varies only
  logarithmically with the accretion rate at super-Eddington
  luminosities.  The jet power follows initially the accretion rate
  but remains a constant after the transition from super- to sub-
  Eddington.  At lower accretion rates at the end of the MAD phase
  the disk becomes thin and the jet may stop altogether. These new
    estimates of the jet power and disk luminosity that do not simply follow
    the mass fallback rate should be taken into account when
    searching for TDEs and analysing light curves of TDE
    candidates. Identification of some of the above mentioned
    transitions may enable us to estimate better TDE parameters. 
\end{abstract}


\section{Introduction}
\label{s.introduction}

The surprising X-ray emission from the  tidal disruption events (TDE)  candidates {\it Swift} J1644 \citep{bloom11,levan11,burrows11} and {\it Swift} J2058 \citep{cenko12} led to a
re-examination of the accretion processes that take place within these events. 
25 years ago, \cite{rees88} outlined the basic
dynamical processes relevant to TDEs. A star  is disrupted by a supermassive black hole. 
The stellar material is spread out and it  returns, after a delay, to 
the vicinity of the black hole where it  forms an accretion disk. If the disrupted stellar 
material has a uniform distribution of orbital binding energy per unit
mass\footnote{This assumption that was verified by various numerical simulations, see e.g. \cite{lkp09,guillochon+13}}, 
the infall rate  onto the accretion disk satisfies: $dM/dt \propto t^{-5/3}$ \citep{phinney89}.
At the peak accretion rate, the luminosity would be
super-Eddington \citep{1999ApJ...514..180U}. Within this disk, gravitational energy is
dissipated and the heat is radiated in the usual quasi-thermal
fashion. If it is thermally radiated by the accretion disk, the
associated temperature would be in the extreme ultra-violet (EUV)
or perhaps the soft X-ray band \citep[see e.g.][]{lodato11}. Consequently, most searches
for such events hitherto have been carried out in the EUV region. Indeed several such candidates were found  \citep{Gezari+08,Cappelluti+09,vanVelzen+11,Arcavi+14}.

In early 2011, \cite{giannios11} suggested that TDEs would involve ejection of relativistic jet that 
would give rise to a radio signal when it is slowed down by the
surrounding matter. This prediction was verified when the radio
emission from {\it Swift} J1644 was discovered a few months later
\citep{zauderer11}. However, it turned out that the jet is also a
powerful source of X-rays \citep{bloom11,burrows11}. These X-rays  arise, most likely, from the
inner part of the jet, probably due to internal shocks taking place
there. A few months later, \cite{cenko12}
discovered a similar X-ray signature from another TDE candidate  ({\it
  Swift} J2058). The latter source also showed an optical emission,
most likely  the thermal component arising from the accretion
disk \citep{cenko12}.

The interplay between the (unexpected) nonthermal jet emission and the thermal disk component is intriguing. 
Over a short period of a few months the accretion rate spans a large range of values, beginning at super-Eddington, turning to sub-Eddington power and  then diminishing. \cite{krolik12} addressed this issue  using   simplified models for both the thermal luminosity and for the jet power. 
For the former they assumed that the luminosity is capped at the
Eddington luminosity at the super-Eddington phase and then it
decreases following the mass accretion rate. For the jet power
\cite{krolik12} related  the Blandford-Znajek   \citep[hereafter BZ,
][]{blandfordznajek-77,mck05,hk06} jet power to the pressure at the
inner parts of the disk. This pressure is linearly related to the mass
accretion rate at the super-Eddington phase.   But the pressure is independent of the mass accretion rate during the radiation dominated phase,  that follows the super-Eddington phase \citep{moderski96}.   This led \cite{krolik12} to suggest that the jet luminosity will be   constant  during this phase. At sufficiently low accretion rates the disk becomes gas pressure dominated and the pressure will decrease as the accretion rate decreases. Thus one would expect two transitions  in the jet's  and disk's light curves. 
One when the accretion rate drops below Eddington and the other when
the disk becomes gas pressure dominated. A third transition would
arise at very low accretion rate, once the disk becomes an advection
dominated accretion flow (ADAF)
\citep{abra88,1995ApJ...452..710N}.

Our goal here is to confront these rather simple estimates with a more
detailed computation of the accretion disk structure as well as with
recent numerical results on the jet luminosity.
To this, end we use relativistic slim disk
solution \citep{sadowski.phd} to estimate the accretion disk
structure over a large range of accretion rates. The slim disk model
generalises the standard thin disks \citep{nt73} towards
high accretion rates. It allows for non-Keplerian rotation, radial
gradients of pressure, and advection of heat. As a result, when the accretion 
rate is super-Eddington, the disks are no longer radiatively efficient and 
their angular momentum profile can significantly depart from Keplerian.
The slim disk model uses $\alpha$ viscosity and assumes the accretion rate
is constant, i.e., there are no outflows.

Simulations of jet formation from accreting BHs have led to different
estimates of the jet luminosity.
The early, pioneering simulations of jet formation in general
relativistic magnetohydrodynamic (GRMHD) simulations
\citep{2003ApJ...592.1060D,hk06} suggested that the magnetic pressure
in the funnel regions of the jet is determined by the thermal pressure
of the accretion disk near the BH horizon. It also appeared that the
accretion disk structure did not seem to be affected by the presence
of the jets or by how strong the jets are \citep{beckwith08}. 
Later, it became clear that both disk thermal pressure and jet magnetic pressure can be
much higher than previously simulated in GR and can be as high as the
ram pressure of the infalling gas \citep{tchekh11,tchekh+12,mtb12}. When the
jet pressure reaches this limit, the magnetic field is strong enough
to obstruct the accretion of gas onto the BH, and this leads to the
formation of a \emph{magnetically arrested disk}, or a MAD
(\citealt{tchekh11}; see \citealt{Igumenshchev+03,Igumenshchev08} for
simulations of MADs in Newtonian relativity and
\citealt{Bisnovatyi+74,Bisnovatyi+76,Narayan+03,tchekh15} for an
analytic consideration).
In the following we will combine the simpler jet luminosity estimates
based on the disk pressure with the model fits
coming from sophisticated numerical simulations. 
  
We begin in \S
\ref{s.jet_power} examining different methods for estimating the
jet power. In \S \ref{s.disk_model} we discuss the slim disk model
of an accretion flow. We present both the thermal emission and
the jet power and we compare the expected behaviour of the jet and disk
luminosity for different accretion rates in \S
\ref{s.luminosities}. We examine the implications for these findings
to tidal disruption events in \S \ref{s.TDE}.  We conclude in \S \ref{s.discussion} with a
discussion of the limitations of the analysis and possible
observational implications.

\section{Jet power}
\label{s.jet_power}

At the order of magnitude
level, the Poynting luminosity of  a BZ  jet that emerges from the vicinity of a black hole can be estimated through a simple dimensional argument. The local magnetic energy density near the black hole is proportional to   $B^2$, where $B$ is the
poloidal field intensity near the horizon. The jet power depends on this energy density and on the area  from which the jet emerges. This area is of order  $ \gsim \pi r_{g}^2$,  where  $r_g$
is the gravitational radius of the black hole. 
Combined the overall luminosity can be written as  $P_{\rm jet} = f(a/M) ~c~(B^2/8\pi)~ (\pi r_{g}^2) $, 
where M is the black hole's mass and  $a_*\equiv a/M$ its specific angular momentum. The function 
$f(a_*)$ is  dimensionless and we approximate it here as $f(a_*)
\approx a_*^2$ \citep[e.g.,][]{2012JPhCS.372a2040T,tchekh15}. 
The task now is to better estimate the magnetic field ($B$) and the area from which the outflow emerges. 

\subsection{The pressure formula}
\label{s.pjetpre}

The jet is in pressure balance with the inner parts of the accretion
disk, therefore the strength of the magnetic field is determined by the pressure in the
inner disk. \cite{beckwith08} have demonstrated that the
magnetic pressure near the horizon is generally bounded above by the
maximal pressure in the equatorial plane near the inner edge of the
disk, $p_{\rm max}$, and it is bounded below by the magnetic pressure
at that location. Thus, we can estimate the strength of the magnetic
energy density using $p_{\rm max}$, the maximal pressure near the edge
of the disk: \be B^2 /8 \pi = \beta_B~ p_{\rm max}, \ee where the
factor $\beta_B$ is an unknown dimensionless factor of order unity.  A
second factor is the size of the region from which the jet emerges. We
use the radius of the innermost stable circular orbit (ISCO), $R_{\rm
  ISCO}$, to characterise the size of
the inner disk, which is comparable but smaller than the position of
the maximal pressure. Together, we can write the BZ luminosity as: 
\be
\label{e.pjetpre1}
P_{\rm jet,P} \equiv \beta_B \pi R_{\rm ISCO}^2 p_{\rm max} a_*^2 c ~\lsim~ \pi R_{\rm ISCO}^2 p_{\rm max} a_*^2 c  .
\ee
The unknown coefficient $\beta_B$ will be determined in
Section~\ref{s.normalization}
by comparing
with the jet power obtained in numerical simulations.

\subsection{The $H/R$ formula}
\label{s.pjetHR}
 
As in the previous section, we will balance the magnetic pressure
pressure in the jet funnel against the pressure of the accretion disk.
The vertical force balance approximately gives,
\be
\frac{p}{\rho}\approx\Omega_{\rm K}^2 H^2=\frac{V_{\rm K}^2 H^2}{R^2},
\ee
where $p$ and $\rho$, are the midplane
pressure and the corresponding density of the
disk, respectively, $\Omega_{\rm K}$ and $V_{\rm K}$ are the Keplerian
velocities
at radius $R$, and $H$ is
disk half-thickness. Applying the vertical equilibrium at the radius
of the pressure maximum, $R$,
 to eq.~\ref{e.pjetpre1}, we get,
\be
P_{\rm jet,H/R} \equiv \beta'_B \pi V_{\rm K}^2
\Sigma H a_*^2 c,
\ee
where we introduced the surface density $\Sigma=2\rho H$, and 
allow the coefficient $\beta'_B$ to differ from $\beta_B$. The rest mass
conservation
requires, 
\be
\dot M=2 \pi R \Sigma V_{\rm r},
\ee where $\dot M$ is the
accretion rate and $V_{\rm r}$ is the absolute value of the radial velocity. Using this formula we
get,
\be
\label{P.hr1}
P_{\rm jet,H/R} = 
\frac{\beta'_B}{2} \dot M c^2\frac{V_{\rm K}^2}{V_{\rm r} c}  \frac{H}{R} a_*^2.
\ee
For radiatively inefficient accretion flows, the magnitude of the radial velocity $V_{\rm r}$
is comparable to the Keplerian velocity $V_{\rm K}$ and does not
depend on the accretion rate. We may therefore approximate in this
case eq.~\ref{P.hr1} and write,
\be
\label{P.hr2}
P_{\rm jet,H/R} \propto \dot M c^2 \frac{H}{R} a_*^2.
\ee

\subsection{Normalisation}
\label{s.normalization}

Numerical simulations of jets in both optically thin and thick,
radiatively inefficient MAD disks \citep{tchekh11,tchekh+12,mtb12,mckinney+harmrad}
provide the missing scaling factor and
give \citep{tchekh15},
\be
\label{e.pjetHRriaf}
P_{\rm jet,H/R,RIAF} = 1.3\, \dot M c^2 \frac{H/R}{0.3} a_*^2.
\ee
Because of the taken assumptions, this formula for the jet power is valid
only for radiatively inefficient disks --- ADAFs or super-critical
(or super-Eddington)
disks, i.e., in the limit of lowest and highest accretion rates.
 To estimate the power of the jet in disks with not so extreme
accretion rates, one has to use more general formulation
(eq.~\ref{e.pjetpre1} or eq.~\ref{P.hr1}).

All the three formulae are expected to give the same estimate of the
jet power. To satisfy this condition, we choose the coefficients
$\beta_B$ (eq.~\ref{e.pjetpre1}) and $\beta'_B$ (eq.~\ref{P.hr1}) so
that the jet power estimates agree with eq.~\ref{e.pjetHRriaf} for $\dot M
\gg \Medd$, and finally get,
\be
\label{e.pjetpre}
P_{\rm jet,P} = 4.0\, \pi R_{\rm ISCO}^2 p_{\rm max} a_*^2 c,
\ee
and
\be
\label{e.pjetHR}
P_{\rm jet,H/R} = 0.16\, \dot M c^2\frac{V_{\rm K}^2}{V_{\rm r} c}  \frac{H/R}{0.3} a_*^2.
\ee

\section{Disk model}
\label{s.disk_model}

To model an accretion disk we use the general relativistic slim disk solutions
of \cite{sadowski.phd}. The slim disk model generalises the standard thin
disk \citep{ss73,nt73} to arbitrary accretion rates. It allows for non-Keplerian
rotation and advective cooling by photons trapped in the flow. It assumes
constant accretion rate, i.e., it does not allow for outflows, and it adopts
the $\alpha$ prescription for viscosity. The slim disks 
reduce in the limit of small accretion rates to the standard thin,
relativistic, Keplerian
disk. 

The system has a characteristic luminosity,  the Eddington luminosity, $L_{\rm Edd}$:
\be
\label{e.ledd}
L_{\rm Edd}=\frac{4\pi GMc}{\kappa_{\rm es}}=1.25\times 10^{38} \frac{M}{M_\odot} \,\rm erg/s. 
\ee
This corresponds to  the Eddington accretion rate, here defined as:
\be
\label{e.mdotedd}
\dot M_{\rm  Edd}=\frac1\eta\frac{L_{\rm Edd}}{c^2}=\frac1\eta\frac{4\pi GM}{c\kappa_{\rm es}}=
2.44\times 10^{18} \frac{M}{M_{\odot}} \,\rm g/s,
\ee 
where we put the efficiency of a thin disk around a non-rotating BH, $\eta=0.057$,
and $\kappa_{\rm es}=0.4\, \rm cm^2/g$. 
Once the accretion rate is near and above $\dot M_{\rm  Edd}$, photons do not have enough time to 
diffuse out of the disk and a fraction of them is advected with the flow.
This extra advective cooling modifies the structure of the disk. In particular, the 
disk radiates  less efficiently and its luminosity scales as \citep{1980AcA....30..347P},
\be
\label{eq.llog}
L\approx L_{\rm Edd}\left(1+\log \frac{\dot M}{\dot M_{\rm Edd}}\right). 
\ee

\begin{figure}
  \centering
\includegraphics[height=1.075\columnwidth,angle=270]{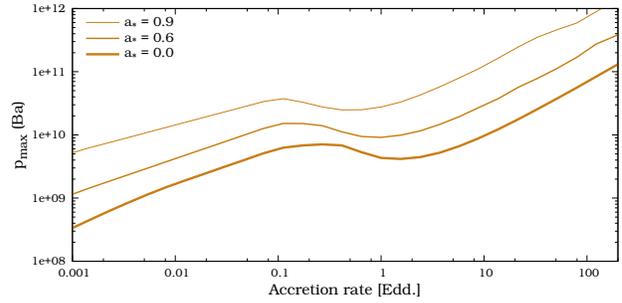}
\caption{The maximal pressure in the disk as a function of the accretion rate for BH spins $a_*=0.3$, $0.6$ and $0.9$ and $M=10^7M_\odot$.}
  \label{f.Pmaxvsmdot}
\end{figure}

The power of the jet depends on the parameters of the underlying accretion flow.
The pressure formula for the jet power (eq.~\ref{e.pjetpre}) is parametrised
in terms of the maximal total pressure in the equatorial plane, $p_{\rm max}$. Fig.~\ref{f.Pmaxvsmdot}
presents $p_{\rm max}$ as a function of the accretion rate for three values of 
BH spin: $a_*=0.0$ (thickest), $0.6$, and $0.9$ (thinnest line), assuming a
BH mass $M=10^7M_\odot$. For the lowest accretion rates, $\dot M\lesssim0.1 \Medd$,
the disk is gas pressure dominated and the pressure at 
fixed radius is expected to follow $\dot M^{4/5}$ \citep{ss73}. However, the radius of the pressure
maximum is not fixed and the profiles of this quantity follow
this dependence only qualitatively. When accretion rate exceeds
$\sim 0.1 \Medd$ the disk is radiation pressure dominated. The standard thin disk
theory predicts that the pressure at the equatorial plane at 
a given radius is independent of the accretion rate. This explains the flattening
of $p_{\rm max}$ profiles around $0.1 \Medd$. Because the pressure maximum is not 
at a fixed radius and because the advective cooling gradually kicks in when approaching
$\Medd$, the $p_{\rm max}$ slightly varies with $\dot M$. For super-critical (exceeding
$\Medd$) accretion rates, the disk enters the slim disk regime and the 
maximal pressure is proportional to the accretion rate.

\begin{figure}
  \centering
\subfigure{\includegraphics[height=1.075\columnwidth,angle=270]{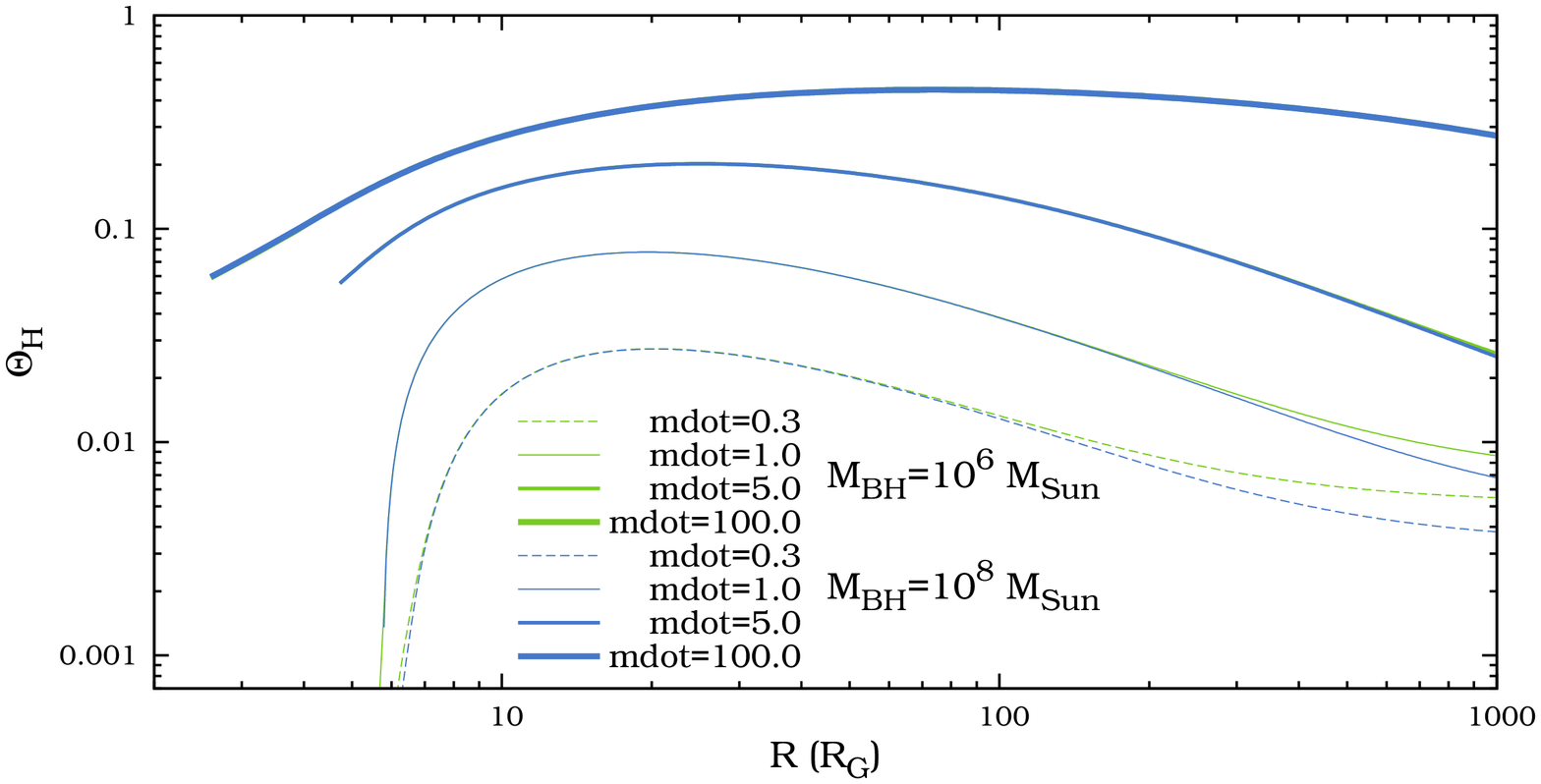}}
\subfigure{\includegraphics[height=1.075\columnwidth,angle=270]{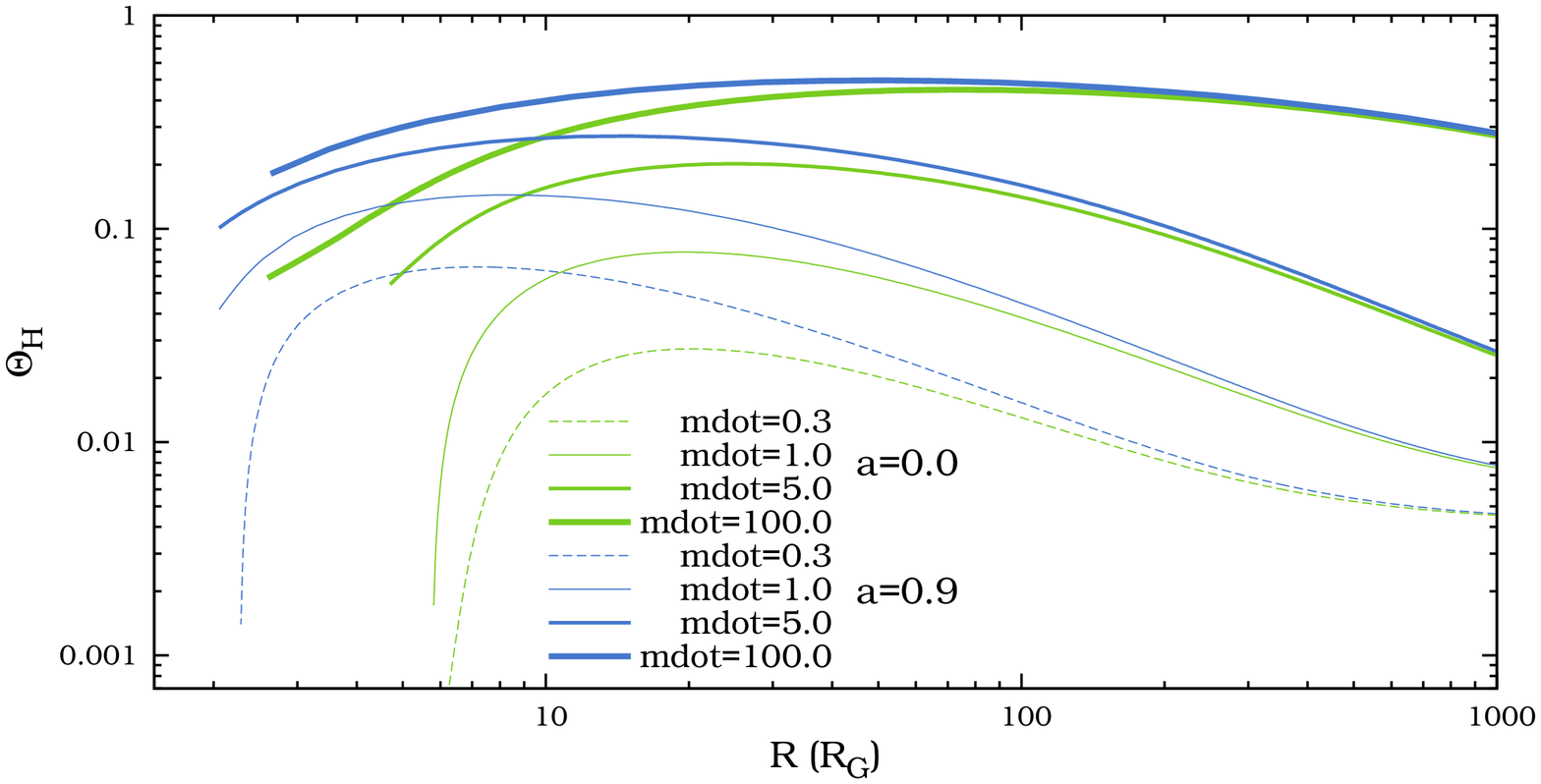}}
\caption{Radial profiles of disk thickness (defined as the opening angle $\Theta_{\rm H}={\rm atan}(H/R)$) for various accretion rates. 
The top panel compares two BH masses ($10^6$ vs $10^8M_\odot$)
assuming zero BH spin, while the bottom
one compares two values of BH spin ($a_*=0$ vs $0.9$) for a BH mass $10^7M_\odot$.}
  \label{f.HRn}
\end{figure}

The $H/R$-based estimates of the jet power (eqs.~\ref{e.pjetHRriaf} and
\ref{e.pjetHR}) depend on the disk
thickness which we parametrise by the maximal disk opening angle, $\Theta_{\rm H}={\rm arctan}(H/R)$.
Fig.~\ref{f.HRn} presents profiles of disk thickness for various accretion rates,
BH spins and masses. The top panel corresponds to a non-rotating BH. 
For $\dot m = \dot M/\dot M_{\rm Edd}=0.3$ the thickest radiation pressure dominated
region is located between $R=10$ and $100$. Disk thickness is determined
there by the local radiative flux. Because both the flux and the vertical
component of the gravity are proportional to the BH mass, the disk opening
angle does not depend on the BH mass. Further out, where gas pressure dominates,
the BH mass has a slight impact on disk thickness. The extent of the radiation pressure
dominated region increases with accretion rate and reaches $R\gtrsim
1000$ for highly super-critical accretion rates.

The bottom panel of Fig.~\ref{f.HRn}  depicts  the disk opening angle for a fixed BH mass $M=10^7 M_\odot$ 
and two values of BH spin, $a_*=0.0$ (green) and $0.9$ (blue lines). The disks
(and hence the thickness profiles) extend more inward for the rotating BH. This  reflects the fact that
the radius of the innermost stable circular orbit decreases with increasing BH spin.
At the same time, for a fixed accretion rate, disks around rotating BH
have a higher maximal thickness than their non-spinning counterparts. This reflects the
higher accretion  efficiency that  results in a higher luminosity.

\begin{figure}
  \centering
\includegraphics[height=1.075\columnwidth,angle=270]{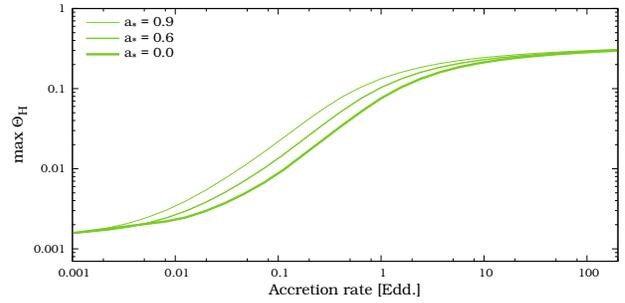}
\caption{The maximal opening angle, $\Theta_{\rm H}$, as a function of the accretion rate for BH spins $a_*=0.3$, $0.6$ and $0.9$.}
  \label{f.HRvsmdot}
\end{figure}

Finally, in Fig.~\ref{f.HRvsmdot} we plot the maximal disk opening angle as
a function of accretion rate for three BH spins. When the disk is gas pressure
dominated at all radii ($\dot M\lesssim 0.001\Medd$) the maximal  disk
thickness is located far from the BH and it does not depend on its spin ---
the lines therefore coincide. In the intermediate, radiation pressure dominated,
regime the disk thickness increases with the BH spin, as discussed above. For
the radiatively inefficient (slim) regime, the maximal disk thickness 
saturates at $\Theta_{\rm H}\approx0.3$. This reflects the fact that the thickness
of advection dominated accretion flows depends only on the  ratio of the radiation to gas
pressure, and approaches the limiting value $\Theta_{\rm H}\approx0.3$ for radiation pressure dominated super-critical disks \citep{narayanyi94,vieira+15}.

\section{Jet and disk luminosities}
\label{s.luminosities}

\begin{figure}
  \centering
\includegraphics[height=1.075\columnwidth,angle=270]{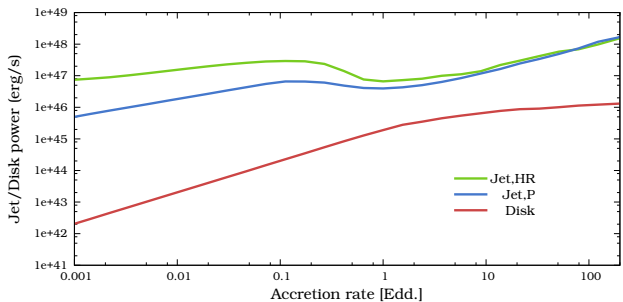}
\caption{ Jet (green and blue) and disk (red line) luminosities for $M_{\rm BH}=10^7 M_\odot$ and
  BH spin $a_*=0.6$ a function of accretion rate.  The
  green line corresponds to eq.~\ref{e.pjetHR}. The blue line shows the
 jet power calculated according to  eq.~\ref{e.pjetpre}, which we take below
 as our fundamental jet power estimate. 
}
  \label{f.Ljetsdisk}
\end{figure}

The standard model of thin disks \citep{ss73} predicts that energy
liberated is proportional to the mass accretion, $\dot M$, and
determined by the accretion efficiency, $\eta$, \be L=\eta \dot M c^2.
\ee For non-rotating BH $\eta=0.057$. As discussed above, once the
accretion rate approaches and exceeds the Eddington limit, the
efficiency is decreased. These facts are reflected in the disk
luminosity profile shown in Fig.~\ref{f.Ljetsdisk}, corresponding to
a BH spin $a_*=0.6$ and a BH mass $M_{\rm BH}=10^7M_\odot$, with the red
line. As long as $\dot M<\dot M_{\rm Edd}$, the disk luminosity
increases linearly with the accretion rate. Once this limit has been
exceeded, the luminosity grows roughly with logarithm of $\dot M$
(eq.~\ref{eq.llog}).

We now apply the jet power formulae derived in Section~\ref{s.jet_power}
and estimate the jet power for each accretion rate using the
corresponding slim disk solution as the underlying disk model.

The pressure formula (eq.~\ref{e.pjetpre}) reflects the fact that the
disk and jet pressures balance each other. Knowing the disk thermal
pressure we may therefore estimate the jet magnetic pressure, and
therefore the magnetic flux in the jet. This quantity, together with
the known BH spin, provides an estimate of the jet power. Its
dependence on mass accretion rate is shown in Fig.~\ref{f.Ljetsdisk}
with the blue line. For a fixed BH mass and spin eq.~\ref{e.pjetpre}
depends only on the maximal value of the disk thermal pressure, and
therefore the profile of the jet power estimated this way resembles
the profile of the corresponding maximal disk pressure
(Fig.~\ref{f.Pmaxvsmdot}). The jet power in the super-Eddington regime
grows proportionally to the accretion rate, remains roughly constant in
the radiation-pressure dominated thin disk regime ($0.1<\dot M/\dot
M_{\rm Edd}<1$), and follows $\dot M^{4/5}$ for gas-pressure
dominated thin disks. These properties result in very powerful jet power
estimates (exceeding the disk luminosity by 2-3 orders of magnitude) for
the thinnest disks. This is inconsistent with observations
\citep[see, e.g.,][]{2004MNRAS.355.1105F,2011ApJ...739L..19R} and suggests
that the assumptions behind the pressure formula break down in this regime.

The other formula for the jet power (the $H/R$ formula, eq.~\ref{e.pjetHR}), 
similarly uses the disk thermal pressure as the proxy for
the jet magnetic pressure and magnetic flux at the horizon. However,
the disk pressure is replaced with the disk thickness using the
vertical equilibrium equation, and the radial velocity $V_{\rm r}$ and
the disk thickness $H$ are introduced.
These are taken directly from the numerical solutions of slim disks.
The corresponding jet power is
plotted with the green line in Fig.~\ref{f.Ljetsdisk}. It
coincides with the pressure formula for $\dot M>\dot M_{\rm Edd}$ and
stays close for lower accretion rates.  This fact proves that both
formulae properly identify the disk thermal pressure, although the 
$H/R$ formula does it indirectly. From now on we will not distinguish between
these two ways of estimating the jet power and use eq.~\ref{e.pjetpre}
as the proxy.

\begin{figure}
  \centering
\includegraphics[height=1.075\columnwidth,angle=270]{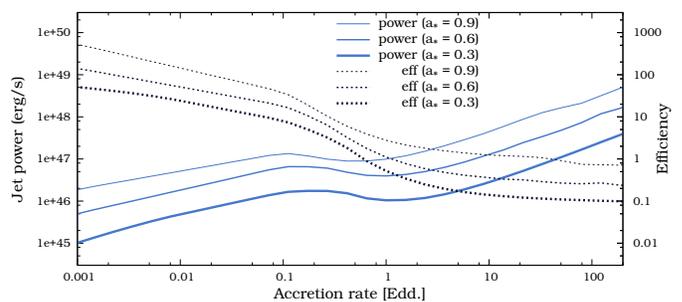}
\caption{ Jet power (blue lines, left $y-$axis) and its efficiency (given by
  eq.~\eqref{eq.eff} and shown with dotted black
  lines, right $y-$axis) for $M_{\rm BH}=10^7 M_\odot$ and BH spins $a_*=0.3$, $0.6$ and $0.9$. }
  \label{f.Ljetvsmdot}
\end{figure}

In Fig.~\ref{f.Ljetvsmdot} we plot the jet power and its efficiency,
$$
\label{eq.eff}
\eta=\frac{L_{\rm jet}}{\dot M c^2},
$$
as a function of accretion rate for three values of BH spin with solid
blue and dotted black lines, respectively. The profiles
of the jet power for all spins agrees with the discussion above. The jet
power is expected to grow roughly with BH spin squared, therefore, the
higher the spin, the larger is the jet power. 
In the limit of highest accretion rates, the efficiency calculated
according to both formulae for spin $a_*=0.9$ is close to
$\eta=1$, what reflects the fact that the efficiencies of jets
observed in GRMHD simulations may be very high, and even exceed $1$
for the highest spins and magnetically arrested disks
\citep{tchekh11,tchekh+12}.

The jet efficiency remains roughly constant as long as we are in the
radiatively inefficient regime ($\dot M>\dot M_{\rm Edd}$). Below that
limit the efficiency increases with decreasing accretion rate. Because the disk
thermal pressure stays roughly constant
in the radiation-pressure dominated, thin disk regime, when the
accretion rate changes by an order of magnitude, or more, the
efficiency must increase by that amount. Furthermore, the pressure
dependence on accretion rate for gas-pressure dominated disks is
$P\propto \dot M^{4/5}$, what further increases the efficiency. Such
extreme efficiencies for the thinnest disks ($\eta \approx 100$ for $0.001 \dot M_{\rm Edd}$)
are unphysical and inconsistent with the lack of observed jets in 
high/soft state of Galactic X-ray binaries.

Which assumptions we took break down in the sub-Eddington regime?  Why
are the jets not there? It is probably the assumption that $\beta_B$
that determines the ratio of the magnetic pressure to the disk
pressure remains constant and is independent of the dramatic changes
that take place in the disk when it transits from a thick to thin disk
and from radiation pressure to gas pressure dominated regime. In
particular, one expects the radial disk velocity to decrease and the
large-scale magnetic flux to diffuse outward faster than it is being
advected inward by the accretion flow
(\citealt{1994MNRAS.267..235L,2012MNRAS.424.2097G,2013MNRAS.430..822G};
see however \citealt{2008ApJ...677.1221R}).  In addition, one expects
for thin disks a larger gap between the inner part of the disk and the
black hole. As the accretion rate decreases, less and less material is
within this gap and the thin disk may not be able to drag the
necessary magnetic field to the vicinity of the black hole.

\section{Jet and disk in tidal disruption events.}
\label{s.TDE}

We turn now to apply the previous result to a TDE. We consider a star
with mass  $M_*$ (measured in solar masses) that is disrupted by a massive black hole (BH).
We  approximate the main
sequence mass-radius relation by $R_* \approx R_\odot M_*^{(1-\xi)}$;
$\xi \simeq 0.2$ for $0.1 < M_* \le 1$, but increases to $\simeq 0.4$ for
$1 < M_* < 10$ \citep{kipp94}. Finally we define $k$ as the apsidal motion constant (determined by the star's radial density
profile) and $f$ is its binding energy in units of $GM_*^2/R_*$ \citep{phinney89}.
In numerical estimates, we scale $k/f$
to the value for fully radiative stars, $0.02$,  because this is a reasonable approximation
for main sequence stars with $0.4 M_{\odot} < M < 10 M_{\odot}$
\citep{kipp94}. 

Assuming that the 
mass accretion rate follows the fall back time of stellar material
onto the central black hole, and that the disrupted star has a uniform distribution in orbital binding
energy per unit mass, matter returns to the region near the pericenter
radius
at a rate $\dot M \propto (t/t_0)^{-5/3}$ \citep{phinney89}.  The characteristic timescale
$t_0$ for initiation of this power-law accretion rate is the orbital period for the
most bound matter \citep{lkp09,krolik12}:
\begin{equation}\label{eqn:porbamin}
t_0 \simeq 15 \times 10^5 M_*^{(1-3\xi)/2}
   M_{BH,7}^{1/2}\left(\frac{k/f}{0.02} \right)^{1/2} \hbox{~s}.
\end{equation}
  Using this time
scale we calibrate the maximal accretion rate\footnote{See
    however \cite{Shiokawa+15} for caution concerning the onset of
    accretion in TDEs and the possibility of a lower maximal accretion
    rate that takes place at a later moment.} as:
\begin{equation}\label{eq:charlum}
\dot M_{\rm peak}\sim 0.3 \times 10^{27}  M_*^{(1+3\xi)/2}
    M_{BH,7}^{-1/2}(k/f)^{-1/2}\rm~g~s^{-1}.
\end{equation}

\begin{figure}
  \centering
\includegraphics[height=1.075\columnwidth,angle=270]{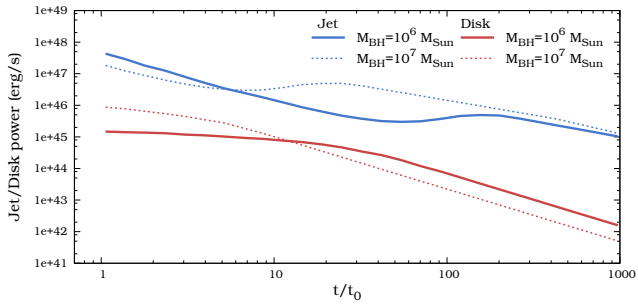}
\caption{Disk luminosity (red) and jet power (blue line) as
  a function of time for a tidal disruption event of a $M_*=1$ star by
a $M_{\rm BH}=10^6M_{\odot}$ (solid) or $10^7M_{\odot}$ (dotted lines)
BH. Other parameters: $M_*=1$,
  $k/f=0.02$, and $\xi=0.2$. }
  \label{f.Ljetvst}
\end{figure}

We convolve now the previous estimates of jet luminosity and disk
power with the accretion rate evolution to obtain the expected light
curves. These are shown for $10^6M_\odot$ and $10^7M_\odot$ BHs
with $a_*=0.6$ spin in Fig.~\ref{f.Ljetvst}. A quick inspection 
reveals that, as expected, even with the more detailed calculations,
the thermal disk light curves (red lines) follow more or less the simple estimates
of \cite{krolik12}. The disk luminosity is roughly constant at short
time scales (when $\dot M > \Medd$). After the transition to $\dot M <
\Medd$, which takes place roughly at $t/t_0 = 6$ and $20$ for
$10^7M_\odot$ and $10^6M_\odot$ BHs, respectively, the
disk luminosity decreases proportionally to the mass
accretion rate. For a higher BH mass (dotted lines) the qualitative behaviour
remains the same. However, because the initial accretion rate
(eq.~\ref{eq:charlum}) is lower, the disk enters earlier the
  sub-Eddington and thin disk phases.

The jet luminosity evolution is also consistent with predictions of
\cite{krolik12}. Initially, as long as the disk is
super-Eddington,
the jet power (denoted with blue lines) decreases proportionally to the accretion rate. Once it
enters the sub-critical, radiatively efficient regime, the jet power
hardly changes reflecting the fact that the disk pressure is not
sensitive to the accretion rate. This changes once the disk becomes
gas pressure dominated. Our simplistic formulae predict further
decrease of the jet power, although not as steep as in the
super-critical regime. However, as discussed earlier, thin disks are
not likely to sustain strong magnetic flux at the BH and therefore
should produce less efficient jets than we predict.

\section{Discussion and Conclusions}
\label{s.discussion}

We have used numerical modelling to estimate the jet power and the
accretion disk luminosity as a function of an accretion rate over a
very wide range of accretion rates. Our accretion disk model use the
relativistic slim disk solutions of \cite{sadowski.phd}. It switches
from a thick disk for accretion rates above Eddington to a thin disk
below.  We have used two different expressions for the jet power, one
based on the maximal pressure and another, motivated by numerical
simulation, based on the $H/R$ ratio. Remarkably,
both formulae give comparable results.

Our results confirm earlier expectations \citep{krolik12} that were
based on simple scaling arguments. According to these expectations
the jet power and the disk luminosity do not simply follow the
accretion rate. Specifically, for super-Eddington accretion 
the
jet power follows the accretion rate but the disk bolometric
luminosity varies only logarithmically with it. At sub-Eddington
accretion the jet power becomes roughly constant while the disk
bolometric luminosity follows the accretion rate.  A puzzling behaviour
in our solution is the appearance of a very large efficiency (jet
power vs. accretion rate) at low accretion luminosity. While puzzling
we note that this is not impossible in principle as the BZ jets gets
its power from the rotational energy of the central black hole and as
such could, in principle at least, operate even with lower accretion
rate (provided that it is surrounded by strong enough magnetic
fields).  However, as we discuss shortly below, it is likely
that the jet power switches off completely in this
low accretion rate regime.

It is interesting to note that in our model the jet power is higher by
at least one order or magnitude than the disk luminosity. This holds
as long as a jet exists for the whole range of accretion rates and
time scales considered. This may not be related directly to the
question whether it is easier or more difficult to detect the jet
emission or the disk emission as those depend also on the spectral
energy distribution of
these two components and on sensitivity and coverage of different
detectors in the relevant spectral regimes.

\begin{figure}
  \centering
\includegraphics[height=1.075\columnwidth,angle=270]{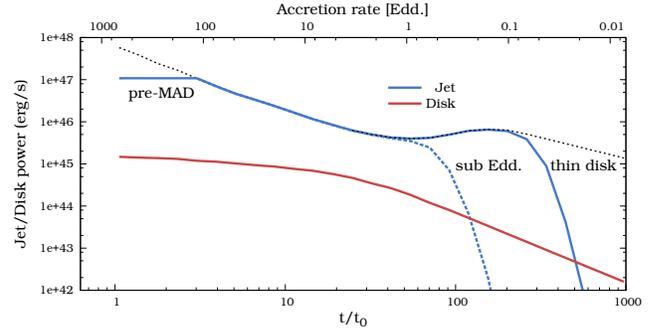}
\caption{
Similar to Fig.~\ref{f.Ljetvst} but only for $M_{\rm
  BH}=10^6M_{\odot}$. The labels above the top axis denote the
accretion rate in Eddington units at given time.
The plateau before $t=3t_0$  schematically
reflects the pre-MAD stage when the accretion rate is too large to form
magnetically arrested disk. The jet power was arbitrarily damped for
$t>200t_0$ to reflect
the fact that the disk enters the thin disk regime ($\dot M<0.1\Medd$)
and is not expected to produce efficient jet at all. The dotted blue
line reflects corresponding decrease in the jet power if the
transition takes place already at $\Medd$.}
  \label{f.Ljetdiskvst}
\end{figure}

It should be stressed that three possibly important effects have not
been taken into consideration in our work.  First, we assumed
(implicitly) that the magnetic fields needed for the BZ mechanism to
operate are there right from the beginning. This might not be the case
in TDEs in which the infalling material is not highly magnetised
initially. \footnote{The
  energetics of Sw J1644 requires the presence of a large-scale magnetic
  flux that exceeds by $3{-}5$ orders of magnitude the magnetic flux
  through a main-sequence star. Such a large magnetic flux could be
  captured by the debris stream from a pre-existing accretion disk
  \citep{tchekh+14}, as recently seen in numerical simulations
  \citep{2014MNRAS.445.3919K}.} Thus it may take some time to build this magnetic field and
in this case we expect a constant (or even increasing) jet power at
early times
\citep{tchekh+14}. This effect was marked in Fig. \ref{f.Ljetdiskvst}
as a solid horizontal line marked ``pre-MAD''.  In other transient
astrophysical source, core-collapse gamma-ray bursts, this stage lasts
for the most duration of the event: in fact, then the MAD onset plausibly marks
the end of the burst \citep{tg15}.

A second potential effect that
we have ignored here is the possibility of the jet switching off when
the disk makes a transition from thick to thin \citep{2013ApJ...767..152Z, tchekh+14}. This transition takes
place shortly after the accretion rate makes the transition from super
to sub-Eddington. As mentioned earlier such a turn off of the jet can
arise from the fact that the weaker accretion rate does not push the
magnetic field sufficiently close to the black hole, or that the
large-scale magnetic flux diffuses outward more effectively than it is
dragged inward by the disk. This possible
effect was marked in Fig. \ref{f.Ljetdiskvst} by arbitrary damping the
jet power for $\dot M<0.1\Medd$ what indicates that the jet
power may not satisfy the simple curve in this region. Note that in
the future, as mass accretion rate drops even further, $\dot M
\lesssim 0.01\Medd$, the accretion disk is expected to transition to a
geometrically-thick radiatively-inefficient ADAF, which can cause
re-launching of the jets and X-ray emission from Sw 1644 \citep{tchekh+14}.

Finally, we assumed that all of the mass fallback ends up reaching the
black hole. The processes of gas circularisation are not
well-understood, and it is possible that the fraction of gas that
reaches the hole depends on the Eddington ratio, and this
can cause additional deviations from simple power-law scalings in time.
Note also that the scaling vs time of intensity in a particular
detector might deviate from a power-law due to the shifting spectrum
of the source coming in and out of the detector bandpass, which might
further complicate the structure of the detected lightcurves.

To conclude, we remark on a few potential observational implications.
First, we note that for all reasonable values of parameters and for all accretion
rates the jet power (as long as the jet exists) is at least one order
of magnitude larger than the disk luminosity (see, e.g.,
Fig.~\ref{f.Ljetsdisk}). This is consistent with observations that 
jet power exceeds accretion disk luminosity in blazars and that the
accretion flow in these systems is in the magnetically-arrested disk regime
\citep{1991Natur.349..138R,2014Natur.510..126Z,2014Natur.515..376G,zdz15}. Moreover, if the jet
is relativistic its radiation would be beamed and enhanced
further. Secondly, we note again that both the jet power (and its
corresponding X-ray emission) and the disk luminosity do not follow in
a simple manner the mass accretion rate.  This should be taken into
account when searching observationally for TDEs or when analysing the
observed light curves of TDE candidates. Particularly interesting is
the possibility \citep{krolik12} that some of the above mentioned
transitions and in particular the transition from super- to sub-
Eddington accretion or the formation of a thin disk could be
identified. This would provide significant new independent information
on the parameters of the TDE and in particular on the masses of
massive black hole. 

\section{Acknowledgements}
We thank Marek Abramowicz for initiating this collaboration. T.P.  was supported by an ERC advanced grant (GRBs) and by the  I-CORE 
Program of the Planning and Budgeting Committee and The Israel Science
Foundation (grant No 1829/12). 
A.S.\ and A.T.\ acknowledge support
for this work 
by NASA through Einstein Postdoctoral Fellowships PF4-150126 and
PF3-140115, respectively, 
awarded by the Chandra X-ray Center, which is operated by the
Smithsonian
Astrophysical Observatory for NASA under contract NAS8-03060. 

\bibliographystyle{mn2e}

\end{document}